\documentclass[fleqn,twoside]{article}
\usepackage{espcrc2}
\usepackage{amsmath}
\usepackage{epsfig}
\usepackage{graphicx}
\usepackage{xspace}
\usepackage[figuresright]{rotating}

\newcommand{\eemm}{\ensuremath{e^+e^-\to\mu^+\mu^-}\xspace}

\newcommand{\eegg}{\ensuremath{e^+e^-\to\gamma\gamma}\xspace}

\newcommand{\dd}{\mathrm{d}}
\newcommand{\vecc}[1]{\mbox{\boldmath $#1$}}

\title{Monte-Carlo Generator Photon Jets for the process \eegg}

\author
{
S.\,I.\,Eidelman$^{a,b}$, 
G.\,V.\,Fedotovich$^{a,b}$
\/\thanks{e-mail: G.V.Fedotovich@inp.nsk.su}
, E.\,A.\,Kuraev $^{c}$,
A.\,L.\,Sibidanov$^{a}$\\
{\it $^a$Budker Institute of Nuclear Physics, 630090, Novosibirsk, Russia},\\
{\it $^b$Novosibirsk State University, 630090, Novosibirsk, Russia},\\
{\it $^c$Bogolubov Lab of Theor. Phys., Dubna, Moscow region, Russia}\\
}

\date{}

\begin{document}

\maketitle

{\large{\bf Abstract}} Monte-Carlo generator with photon jets radiation in collinear 
regions for the process \eegg is described in detail. Radiative
corrections in the first order of $\alpha$ are treated exactly. Large leading
logarithmic corrections coming from collinear regions are taken into account
in all orders of $\alpha$ by applying the Structure Function approach.
Theoretical precision of the cross section with radiative
corrections is estimated to be 0.2\%. This process is
considered as an additional tool to measure luminosity in
forthcoming experiments with the CMD-3 detector at the
$e^+e^-$ collider VEPP-2000.

\section{Introduction}
For the first time the process
$$
e^+(p_+)\ +\ e^-(p_-)\ \to\ \gamma(k_1)\ +\ \gamma(k_2),
$$
\noindent
was considered in the classical papers by L.Brown and
R.Feynman~\cite{feynman}, I.Harris and L.Brown \cite{harris}
in early 1950s and then revised in 1973
by F.Berends and R.Gastmans~\cite{classic}. Due to the large magnitude
of the cross section of this process, it can be exploited
as an independent way to measure luminosity.
Precise determination of luminosity is a key ingredient in
all experiments which study hadronic cross sections at $e^+e^-$ colliders.
As a rule, a systematic error of luminosity measurements represents
one of the largest sources of uncertainty
which can cause  significant reduction of an accuracy of the hadronic cross
sections normalized to luminosity. The process of two-gamma-quantum 
annihilation has essential advantages
for luminosity measurements with respect to those based on events of Bhabha
scattering or annihilation into a muon pair. Indeed, the cross section value
estimated for large angles is of the same order as
that of Bhabha scattering. Events of this process have two
collinear photons at large
angles providing a clean signature for their selection among other
detected particles.
The CLEO-II collaboration was the first to show in practice how the combined
application of the processes $e^+ e^- \to\ e^+ e^- $,
$ \mu^+ \mu^- $ and  $ \gamma\gamma $ helped to
achieve a 1\% accuracy of the luminosity measurement~\cite{cleo}.

\begin{figure}[tbh]
\includegraphics[width=0.45\textwidth]{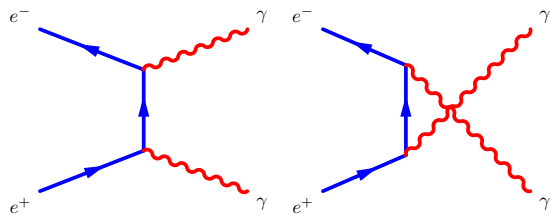}
\caption{\label{fig:gg-fig} Feynman diagrams for 
two-gamma-quantum annihilation.}
\end{figure}

It is worth noting that the dependence of the Born cross
section on the photon polar angle $\theta$ is not so steep as that of Bhabha
scattering and it is symmetric under the transformation
$\theta \to \pi - \theta$ facilitating a study of the
systematic uncertainties on the detector acceptance. In
addition, it is free of difficulties related to both radiation and
Coulomb interaction of the final state particles. It is also of utmost
importance that corresponding Feynman graphs do not contain photon propagators
affected by vacuum polarization effects, Fig.~\ref{fig:gg-fig}.

Therefore, knowledge of this cross section with radiative corrections (RC)
at the level of per mill accuracy is
urgently needed. On the other hand, it is a purely quantum
electrodynamics (QED) process giving
large background while studying hadronic processes
with neutral particles in the final state. These reasons are the main
motivation to consider this process with precise radiative
corrections  and create a Monte-Carlo generator to simulate \eegg events.

\section{Cross section  of the process \eegg with the first order $\alpha$ 
corrections}

The differential Born cross section for the two-gamma-quantum
annihilation in the Born approximation reads

\begin{eqnarray}
\hspace{-0.0 cm} \frac{\dd\sigma_B}{\dd\Omega_1} & = & 
\frac{\alpha^2}{s\beta}\biggl[
\frac{1+\beta^2 c^2_1}{1-\beta^2 c^2_1}
\nonumber \\
 & + & 2\beta^2(1-\beta^2)\frac{1-c^2_1}{(1-\beta^2 c^2_1)^2}\biggr],
\label{Born}
\end{eqnarray}

\noindent where $s=(p_+ + p_-)^2 = 4 \varepsilon^2$, $\varepsilon$ is 
the beam energy and $\beta = v/c = \sqrt{1 - m_{e}^2 /\varepsilon^2}$,
$c_{1} = \cos\theta_1$, $\theta_1=\widehat {\vecc{k_1} \vecc{p_-}}$.
It is assumed that both final photons are detected
and their polar angles with respect to the beam directions are
not too small ($\theta_{1,2}\gg m_e/\varepsilon$).

Following the well known results obtained in~\cite{eegg} and
considering the RC due to emission of virtual and soft real photons,
this part of the cross section can be written as

\begin{eqnarray}
&& \hspace{-1.2 cm} \dd \sigma_{B+S+V}=\dd \sigma_B \biggl\{1 + 
\frac{\alpha}{\pi}
\biggl[(L-1)\biggl(2\ln\frac{\Delta\varepsilon}{\varepsilon} + 
\frac{3}{2}\biggr)
\nonumber \\ 
&& \hspace{-1.2 cm} + K_{SV}\biggr]\biggr\}, 
\quad K_{SV} = \frac{\pi^2}{3}+\frac{1-c^2_{1}}{2(1+c^2_{1})}\biggl[\biggl(1 + 
\nonumber \\
&& \hspace{-1.2 cm} + \frac{3}{2} \frac{1+c_{1}}{1-c_{1}}\biggr)
\ln\frac{1-c_{1}}{2} 
 + \biggl(1+\frac{1-c_{1}}{1+c_{1}} + \frac{1}{2}\frac{1+c_{1}}{1-c_{1}}\biggr)
\nonumber \\
&&\hspace{-1.2 cm} \cdot \ln^2\frac{1-c_{1}}{2} + (c_{1} \to -c_{1})\biggr], 
\end{eqnarray}

\noindent
where $L=\ln(s/m_e^2)$ is a large logarithm. For $s \sim 1~GeV^2$, 
$L\approx 15$. The energy of the radiated soft photons is
implied to be sufficiently low so that they are not detected and its 
value does not exceed some small quantity 
$\Delta \varepsilon \ll \varepsilon$.

Consider now the process of three-gamma-quantum annihilation which can be
treated as a radiative correction for two-gamma-quantum annihilation:
$$ e^+(p_+)\ +\ e^-(p_-)\ \to\ \gamma(k_1)\ +\ \gamma(k_2)\ +\ \gamma(k_3).$$
For the first time the analytic expression for this process
was obtained by M.V.~Terentjev in~\cite{eeggg}. A
much simpler way to obtain the same expression was
suggested in~\cite{chiral-method}, based on the chiral
amplitude method, when all three hard photons are emitted in the so called
non-collinear region (outside narrow cones). The cross section is given by

\begin{eqnarray}
&& \dd\sigma^{e^+e^-\to 3\gamma} = 
\frac{\alpha^3}{8\pi^2 s}\; R_{3\gamma}\,\dd\Gamma\, ,
\\
&& R_{3\gamma} = s\;\frac{\chi_3^2+(\chi_3')^2}{\chi_1\chi_2\chi_1'\chi_2'} - 
\nonumber \\
&& - 2m_e^2\biggl[\frac{\chi_1^2+\chi_2^2}{\chi_1\chi_2(\chi_3')^2}
+ \frac{(\chi_1')^2+(\chi_2')^2}{\chi_1'\chi_2'\chi_3^2}\biggr] 
\nonumber \\
&& + \mbox{two cyclic permutations},
\nonumber \\
&& \dd \Gamma = \frac{\dd^3k_1\dd^3k_2\dd^3k_3}{k_1^0k_2^0k_3^0}
\delta^{(4)}(p_+ + p_- - 
\nonumber \\
&& - k_1 - k_2 - k_3), \quad 
\dd \Gamma = \frac{s}{2}\frac{\dd \Omega_{1}\dd \Omega_{\gamma}
x_{3}\dd x_{3}}{[2 - x_{3}(1-c_{3})]}, \nonumber 
\end{eqnarray}
\noindent where $\chi_i=k_ip_-$, $\chi_i'=k_ip_+,$ i=1,2,3,
$x_{i} = k^0_{i}/\varepsilon,$ $c_{i} = \cos(\theta_{i}),$  
 $\theta_{i} = \widehat {\vecc{p_-} \vecc{k_i}}. $

Energy-momentum conservation allows to determine kinematics of the final
photons:\\
$ x_1 + x_2 + x_3 = 2,\quad x_1c_1 + x_2c_2 + x_3c_3 = 0 $
\begin{eqnarray} 
&& x_1 = \frac{1-x_3}{1-x_3\sin^2{\frac{\psi}{2}}}, 
\nonumber \\
&& x_2 = \frac{\cos^2{\frac{\psi}{2}}+(1-x_3)^2\sin^2{\frac{\psi}{2}}}
 {1-x_3\sin^2{\frac{\psi}{2}}},  
\nonumber \\
&& c_2 = - \frac{x_1c_1+x_3c_3}{x_2}, \quad \psi = \widehat {\vecc{k_1} 
\vecc{k_3}}. 
\end{eqnarray}
The sum of this cross section with those describing {\it soft} and
{\it virtual} photons radiation does not depend on inner parameters.
It allows to construct a
MC event generator to simulate three-photon events and
to take into account proper selection criteria of a given experiment
as well as include specific detector imperfections.

\subsection{Matching NLO and higher-order (HO) corrections}

It is known that photon jets are radiated in collinear regions along the motion
of electrons and positrons give the dominant
contribution to the cross section. So, in order to achieve a theoretical
precision of about per mill, all enhanced HO corrections must be taken into
account and combined with NLO corrections.
The opening angle of these narrow cones is small and obeys the restrictions:
$1/\gamma \ll \theta_{0} \ll 1$. As a rule, its value is chosen as
$\theta_{0} = 1/\sqrt{\gamma}$. Since the photon radiation
outside these cones is not  enhanced, it is
sufficient to consider only radiation of one photon at large angles~\cite{epjc}
to keep the theoretical accuracy at the per mill level.

For completeness, the cross section with one hard photon emission in the 
collinear
region is presented below. It can be obtained using the method of quasi-real
electrons~\cite{quasi} similarly to~\cite{epjc}.
The theorem of factorization~\cite{factorization} of {\it hard} and 
{\it soft} photons
permits to treat RC in the leading logarithmic approximation in all
orders of perturbation theory in terms of Structure Functions of electron
and positron, ${\cal D}(z,L)$. This fact allows one to write the differential
cross section for the process of two gamma-quantum annihilation integrated
inside the  collinear region as follows:

\begin{eqnarray}
\dd\sigma_{\mathrm{coll}} & = & 
\frac{\alpha}{\pi}\int\limits_{\Delta}^{1}\frac{\dd x}{x}
\biggl[(z+\frac{x^2}{2})\biggl(L-1+\ln\frac{\theta^2_0}{4}\biggr)  
\nonumber \\
 & + & \frac{x^2}{2}\biggr]\left[\dd\tilde\sigma_0(z,1) + 
\dd\tilde\sigma_0(1,z)\right],
\end{eqnarray}
\noindent where $z = 1 - x$ is electron (positron) energy after photon 
emission with energy x.

The {\em shifted\/} Born cross section with reduced energies of the incoming
electrons and positrons has the form
\begin{eqnarray}
&& \hspace{-1.6 cm} \dd\sigma_0(z_1,z_2) =  \frac{2\alpha^2}{s z_1 z_2} 
\cdot   
\nonumber \\
&&  \hspace{-1.6 cm} \cdot \frac{z_1^2(1-c_{1})^2+z_2^2(1+c_{1})^2}
{(1-c^2_{1})(z_1+z_2+(z_2-z_1)c_{1})^2}\dd\Omega_1,
\end{eqnarray}
\noindent
where the scattering angle $\theta_{1}$ is given for the
original c.m. reference frame of the colliding beams, $z_{1}$ and $z_{2}$ are
the energy fractions of electron and positron just before collision
after radiation photon jets. When $z_{1}$ and $z_{2}$ tend to unity,
this cross section is transformed to (\ref{Born}).
One can see that a part of this cross section has a term proportional to
large logarithm  $L=\ln(s/m_e^2)$, due to collinear photon emission,
and it is already contained in Structure Functions.
Therefore, to match NLO and HO corrections and exclude double counting,
the term proportional to large logarithm must be removed.
The remaining non-leading correction referred to as
a {\it compensator} should be
combined with the cross section describing three-photon production to cancel
the dependence on the auxiliary parameter $\theta_{0}$.

\section{Total cross section of the process \eegg+n$\gamma$}

In the following, we summarize the main features of the matching procedure as
implemented in the code MCGPJ~\cite{epjc}. Adding the higher-order RC in the 
leading logarithmic approximation to the complete one-loop result (NLO), 
the {\it master}
formula for the resulting cross section can be represented as follows:

\begin{eqnarray}
&&  \hspace{-1.2 cm} \dd\sigma^{e^+e^-\to\gamma\gamma + (n\gamma)} = 
\int\limits_{0}^{1}\dd z_1
{\cal D}(z_1)\int\limits_{0}^{1}\dd z_2 {\cal D}(z_2)\cdot
\nonumber \\
&&  \hspace{-1.2 cm} 
\cdot\dd\tilde{\sigma}_0(z_1,z_2)\left(1+\frac{\alpha}{\pi}K_{SV}\right)
\Theta(cuts) + 
\nonumber \\
&& \hspace{-1.2 cm} + \frac{\alpha}{\pi}\int\limits_{\Delta}^{1}\frac{\dd x}{x}
\left[\left(z+\frac{x^2}{2}\right)\ln\frac{\theta^2_0}{4}
+\frac{x^2}{2}\right]\cdot 
\nonumber \\
&&  \hspace{-1.2 cm} \cdot\biggl[\dd\tilde{\sigma}_0(z,1) + 
\dd\tilde{\sigma}_0(1,z)\biggr]\Theta(cuts)
\nonumber \\
&&  \hspace{-1.2 cm}
+\ \frac{1}{3}\!\!\!\!\!\!\int\limits_{\stackrel{x_i\geq\Delta}
{\pi-\theta_0\geq\theta_i\geq\theta_0}}\!\! \frac{4\alpha^3}{\pi^2s^2}
\biggl[\frac{x_3^2(1+c_3^2)}{x_1^2x_2^2(1-c_1^2)(1-c_2^2)} + 
\nonumber \\ 
&& \hspace{-1.2 cm} + \mbox{two cyclic permutations}\biggr] \dd \Gamma 
\Theta(cuts),
\end{eqnarray}
\noindent
where ${\cal D}(z)$ is the smoothed representation for the
Structure Functions according to~\cite{strfun}.
A factor $1/3$ in the last term takes into account the identity
of the final photons. The all variables are defined above.

This expression contains the logarithmically
enhanced contributions due to emission of photons at all
powers of $\alpha$ in collinear regions and, as it will be shown
later, provides the cross section accuracy of about $10^{-3}$.
The first term describes
radiation of photon jets  which is approximated
by the convolution of the Structure Functions with the {\it shifted} Born
cross section ($s' = sz_{1}z_{2}$). The step functions $\Theta(cuts)$ stand
for particular cuts applied.
The sum of the last two terms provides cancelation
of the auxiliary parameters $\Delta$ and $\theta_0$.
\begin{figure}[tbh]
    \includegraphics[height=0.99\linewidth]{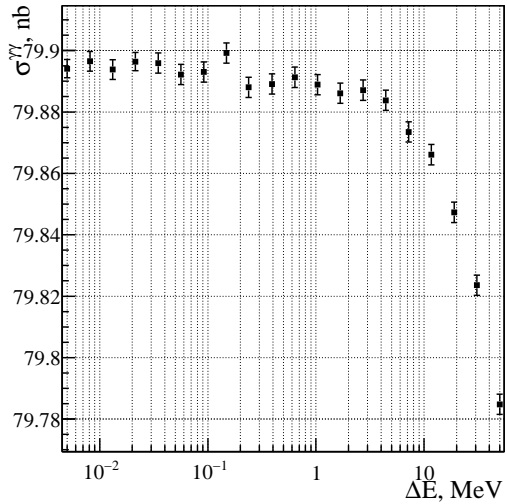}
    \caption{\label{fig:scan_de} Cross section dependence on
    the auxiliary parameter $\Delta\varepsilon$, $\sqrt{s}$ = 900 MeV.}
\end{figure}
A detailed comparison was performed between the results obtained with MCGPJ and
the MC generator based on~\cite{berends} for the cuts modeling
the CMD-2 event selection criteria at the c.m. energy $\sqrt{s}$ = 900 MeV.
These cuts are:
\begin{itemize}
\item The polar angles of the two most energetic photons must be
  inside a range $1.1 < \theta_{1,2}<\pi-1.1$
\item Acollinearity must obey $|\theta_1 + \theta_2-\pi|<0.25$ and
  $||\phi_1-\phi_2|-\pi|<0.15$
\item The energies of the two most energetic photons must be
larger than half of beam energy.
\end{itemize}
\begin{figure}[tbh]
    \includegraphics[height=0.99\linewidth]{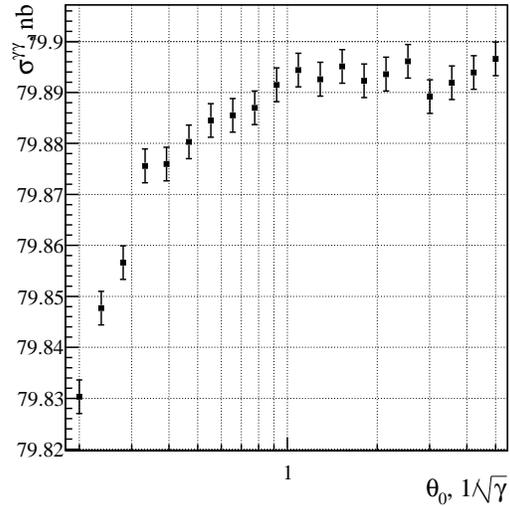}
    \caption{\label{fig:scan_nt0} Cross section dependence on
    the auxiliary parameter $\theta_0$, $\sqrt{s}$ = 900 MeV.}
\end{figure}
In Fig.~\ref{fig:scan_de} the cross section as a function of the
 $\Delta \varepsilon$ is shown when other parameters are fixed according
to selection criteria pointed above. It can be seen that there is
a broad plateau, where the cross
section deviations lie within a band with a width of $\sim 0.2\%$,
whereas  $\Delta \varepsilon$ runs by
more than two orders of magnitude. Only for large values of
$\Delta \varepsilon$ some trend appears which can be explained by
the omitted terms proportional to $\Delta \varepsilon/\varepsilon$.
Similar dependence is seen in
Fig.~\ref{fig:scan_nt0}, where the cross section variations with
the auxiliary parameter $\theta_0$ are presented.
Only for extremely small values of $\theta_0$, when a condition $1/\gamma \ll
\theta_{0}$ is not valid, the cross section falls down by $\sim 0.1\%$ only.
It is an important result, which confirms that
the cross section does not depend on the auxiliary parameters
$\Delta \varepsilon$ and $\theta_{0}$.
\begin{figure}[tbh]
    \includegraphics[height=0.99\linewidth]{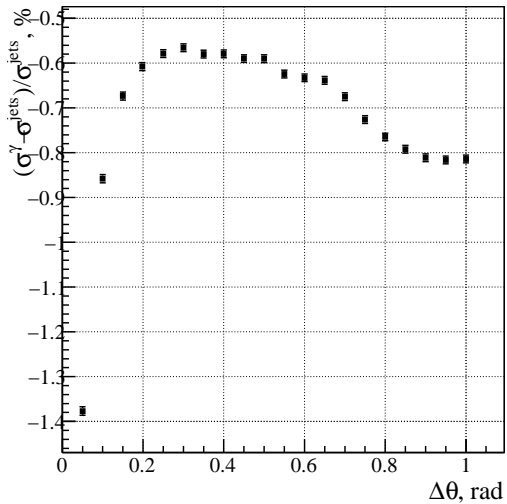}
    \caption{\label{fig:scan_one_jet_th} Relative cross section difference
calculated with the MCGPJ code and generator based on Ref.~\cite{eegg} versus
the angle $|\Delta\theta|$.}
\end{figure}
\begin{figure}[tbh]
    \includegraphics[height=0.99\linewidth]{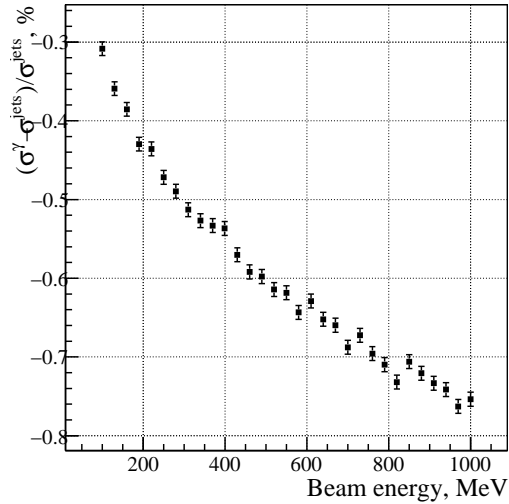}
    \caption{\label{fig:scan_one_jet_en} Relative cross section difference
calculated with the MCGPJ code and generator based on Ref.~\cite{eegg}.}
\end{figure}
It is important to reliably estimate the total theoretical precision
of this cross section with RC. In order to quantify the
theoretical accuracy, an independent comparison has been performed
with the MC event generator based
on~\cite{berends}, where only first-order $\alpha$ corrections are
treated exactly. It was found that the relative difference of the
cross sections is larger than 1\% for small angles $\Delta
\theta<0.1$ radians and it is about $\sim$ 0.6\% for an acollinearity
angle $\sim$ 0.25 radians. The simulation results are presented in
Fig.~\ref{fig:scan_one_jet_th}. It is seen that the difference
practically does not depend on the choice of the value
$\Delta \theta$ with accuracy $\pm 0.1\%$.
The same difference was studied as a function of beam energy when
the acollinearity angle $\Delta \theta$ was fixed to 0.25 radians.
The results are shown in Fig.~\ref{fig:scan_one_jet_en}. In this case
the difference slowly increases with energy from 0.3\% to 0.7\%.
It is an important fact, which means that for this energy band
the radiation of two and
more photons (jets) in the collinear region contributes to the cross
section by amount $\sim$ 0.5\% only. As
the uncertainty of this correction is at least a few times smaller
than the magnitude of the correction itself, we can conclude:
a theoretical precision of the cross section with
RC is certainly better than $\sim$ 0.2\%.
\section{Conclusion}
The main features of the Monte-Carlo generator to simulate events
of the process \eegg are described. The theoretical precision of the
cross section with RC is $\sim 0.2\%$. It was achieved due to the
application of the  Structure Function
approach which allows one to match the enhanced contributions (HO), coming
from the collinear regions, with the cross section containing NLO
corrections. It is proposed to use the generator as a complementary
tool to measure the collider luminosity compared to the regularly used
processes of Bhabha scattering or \eemm.
The precision of collider luminosity determination represents
one of the largest sources of systematic errors which can cause
significant reduction of the accuracy of hadronic cross sections, which are,
 as a rule, normalized to luminosity. The considered cross section
is rather big, events have a clean
signature in the detector and can be easily recognized and selected. From
the theoretical point of view, the main advantage of using the process \eegg
compared to others is the following: the cross section does not contain
the first order of
$\alpha$ corrections due to the vacuum polarization effects and there
is no FSR and Coulomb interaction of particles.

This work is partially supported by INTAS Ref. Nr 05-1000008-8328,
INTAS YSF Ref. Nr 06-1000014-6464 and the Slovak Grant Agency for
Sciences VEGA, Grant Nr 2/7116/27.

\end{document}